\documentclass[aps,twocolumn]{revtex4-1}
\usepackage{amsmath}
\usepackage{amssymb}
\usepackage{bbold}
\usepackage{graphicx}
\usepackage{dcolumn}
\usepackage{bm}
\usepackage{epsfig}
\usepackage{float}
\usepackage[colorlinks=true, citecolor=red, urlcolor=blue, linkcolor=blue]{hyperref}

\usepackage{bookmark}

\renewcommand{\vec}[1]{\boldsymbol{#1}}

\begin{document}

\title{Confining charged particles with time-varying magnetic fields:
toward non-torus configuration of fusion plasmas}

\author{Shao-Wu Yao}
\affiliation{Center of Theoretical Physics, College of Physics, Sichuan University, Chengdu 610065, China}

\author{Bo You}
\affiliation{Center of Theoretical Physics, College of Physics, Sichuan University, Chengdu 610065, China}

\author{Yue-Hao Yin}
\affiliation{Center of Theoretical Physics, College of Physics, Sichuan University, Chengdu 610065, China}

\author{Zhi-Yong Wu}
\affiliation{Center of Theoretical Physics, College of Physics, Sichuan University, Chengdu 610065, China}

\author{Li-Xiang Cen}
\email{lixiangcen@scu.edu.cn}
\affiliation{Center of Theoretical Physics, College of Physics, Sichuan University, Chengdu 610065, China}

\begin{abstract}
We develop protocols to confine charged particles using time-varying magnetic fields
and demonstrate the possible non-torus configuration resulting from the distribution
of single-particle motion orbits.
A two-step strategy is proposed to achieve this goal: preliminary protocols
are contrived by solely considering the magnetic force; afterwards they are
evaluated and selected through numerical solutions to the equation of motion, taking into
account inductive electric fields. It is shown that a fine-tuned tangent-pulse
protocol can maintain its centralized configuration even in the presence of associated
electric fields, which illuminates an alternative approach to designing the confinement
scenario for fusion plasmas.
\end{abstract}

\maketitle

Magnetic confinement is one of the most promising approaches to achieve the
demanding conditions for controlled fusion reactions, e.g., a high temperature
above $100$ million kelvin and stable control of the fusion plasmas
over the characteristic energy-confinement time \cite{boozer2005,ongena2016}.
In the devices of tokamaks
\cite{wesson2004} and stellarators \cite{spitzer1958,helander2014},
steady magnetic-field configurations are exploited with the field lines produced by
the coils lying on the constant-pressure surface of the ideal
plasma torus. The toroidal structure is considered essential for these designs as
it ensures force equilibrium of the confined plasma through a
topologically nested and closed magnetic surface. On the other hand,
rotating magnetic fields (RMFs) \cite{RMF1,RMF2} have been used in compact torus devices,
such as the rotamak \cite{hugrass1980,hugrass1981,jones1998}
and the field-reversed configuration \cite{tuszewski1988,slough1999,slough2000,guo2005,shi2018},
to generate continuous toroidal currents for sustaining the flux and the plasma.
The distribution of magnetic-field lines in these systems is time varying
and exhibits a cyclic opening character.
As the closed field structure seems to be not a necessity anymore,
a further question arises: whether a confinement configuration
beyond the toric geometry is possible by utilizing the RMF or more general
time-varying magnetic fields (TVMFs)?

Theoretical simulation of single-particle motion plays an important
role in the study of magnetic confinement for fusion plasmas \cite{northrop1963,RMP2009}.
Analytical approach has been exploited to reveal stable orbit motion
for the RMF confinement system \cite{hugrass1983,hugrass1987}.
However, owing to the complexity of the equation of motion with respect to
the TVMF itself and its inductive electric field,
one can hardly apply the similar analytical method to explore the confinement
issue other than the RMF system.
Numerical algorithms such as particle-in-cell simulation techniques \cite{birdsall1985,chen2009,bao2014,bao2016}
and single-particle codes \cite{cohen2000,glasser2002,glasser2022} have been widely used
and proven to be powerful tools to the study of the particle trajectory.
In view that these numerical algorithms
rely on the concrete form of internal and external electromagnetic fields
with respect to the confinement configuration, it is a challenge
to seek effective TVMF protocols where the driving field is
yet to be determined.

In this paper we delve into this very issue by a two-step strategy, that is,
we first seek possible TVMF confinement protocols by concerning solely the
magnetic force on the charged particle (the deuteron or the triton),
and check subsequently the validity of the obtained protocols in the presence
of inductive electric fields.
Specifically, we will explore the confinement condition by virtue of
the correspondence relation established between the Newton-Lorentz equation
of the classical particle and the Bloch equation of a class of driven two-level
quantum systems. In the light of rigorous dynamics generated by the quantum
system with respect to geodesic-loop evolution of the Lewis-Riesenfeld (L-R)
invariant \cite{lewis1967,lewis1969}, a couple of TVMF protocols are proposed and
the spatial distribution of the resulting trajectories of the charged particle
is shown to be of
centralized structure. Moreover, numerical calculations display that one of the
proposals, the tangent-pulse protocol with fine-tuned field parameters,
can maintain its confinement capability even including the influence of the
inductive electric field. The result sheds light on a way of exploiting
the TVMF protocol for magnetically-confined fusion scenarios with
non-torus configuration.

Let us start by considering the dynamical evolution generated by a
driven two-level quantum system
\begin{equation}
	H(t)=\frac 12\vec \Omega (t)\cdot \vec \sigma,
\label{hamil}
\end{equation}
where $\vec{\Omega}(t)\equiv \sum_i\Omega_i(t)\hat{e}_i$
stands for a time-dependent external field and $\sigma_i$ ($i = x,y,z$)
denote Pauli matrices satisfying $[\sigma_i,\sigma_j]=2i\epsilon_{ijk}\sigma_k$.
Such a binary (or spin) quantum system is the basic unit of quantum
information processing \cite{nielsen} and its analytical solvability and associated
dynamical control with various driving fields have long been a subject
of general interest \cite{LZ19320,LZ1932,wang1990,nori2010,ding2010,barnes2012,yang2018,li2018,zhao2018}.
Let $|\Psi(t)\rangle$ be the wave function of the two-level
system which satisfies the Schr\"{o}dinger
equation $i\hbar\frac {\partial}{\partial t}|\Psi(t)\rangle =H(t)|\Psi(t)\rangle$.
We invoke the Bloch vector
of the density operator $\rho(t)\equiv|\Psi(t)\rangle \langle \Psi(t)|$, i.e.,
$\vec{\mathcal{R}}(t)=\sum_k \mathcal{R}_k(t)\hat{e}_k$ specified by
$\rho(t)=\frac 12[I_2+\vec{\mathcal{R}}(t)\cdot\vec{\sigma}]$,
and recast the Schr\"{o}dinger equation as
\begin{equation}
\frac {d}{dt}\mathcal{R}_k(t)=\frac 1 \hbar[\vec{\Omega}(t)\times \vec{\mathcal{R}}(t)]_k.
\label{bloch}
\end{equation}
One is led to notice that this Bloch equation has a similar structure with
the Newton-Lorentz equation accounting for the spatial motion of a
charged particle subjected to a pure magnetic force:
$\dot{\vec{v}}(t)=\frac qm\vec{v}(t)\times \vec{B}(t)$, where $q/m$ denotes
the charge-mass ratio and $\vec{B}(t)$ assumes to be a time-varying but
spatially uniform magnetic field.
By setting $\vec{B}(t)=-m\vec{\Omega}(t)/(\hbar q)$
and dividing both sides of the above Newton-Lorentz equation by the constant magnitude
$v=|\vec{v}(t)|$, one obtains
that the orientation of the velocity $\hat{v}(t)\equiv\vec{v}(t)/v$ satisfies
\begin{equation}
\frac {d}{dt}\hat{v}(t)=-\frac 1 \hbar\hat{v}(t)\times \vec{\Omega}(t).
\label{lorentz}
\end{equation}
Consequently, an explicit correspondence $\hat{v}(t)\leftrightarrow \vec{\mathcal{R}}(t)$
is established between the kinematic behavior of the
classical particle and the dynamical evolution of the driven two-level
quantum system.

Complete confinement of
the particle motion indicates that, for an arbitrary initial
state of motion, the average velocity of the particle,
hence the Bloch vector of the counterpart quantum system,
should satisfy
\begin{equation}
\langle \hat{v}_k(t)\rangle_T \equiv \frac 1 T\int_0^T\hat{v}_k(t)dt=0\Leftrightarrow \langle \mathcal{R}_k(t)\rangle_T=0
\label{cond}
\end{equation}
for all three ($k=x, y, z$) directions in the long-time limit $T\rightarrow \infty$.
To delve into the issue and seek TVMF confinement protocols,
we explore the dynamical evolution generated by a sort of
periodically driven quantum models. Specifically, we consider the driving protocol of
the two-level quantum system $H(t)$ that possesses a
dynamical invariant $I(t)$ \cite{lewis1967,lewis1969} evolving along the geodesic curve
in the Bloch space, e.g., the equatorial line described by
\begin{equation}
I(t)=\frac 12[\cos\phi_c(t)\sigma_x+\sin\phi_c(t)\sigma_y],
\label{invar}
\end{equation}
in which $\phi_c(t)$ goes from $0$ to $2\pi$ cyclically.
Following the L-R theory, such a system is exactly solvable
and the general time evolution of the Bloch vector is
given by
$\vec{\mathcal{R}}(t)=\mathcal{U}(t)\vec{\mathcal{R}}(t_0)$
with the evolution operator
\begin{equation}
\mathcal{U}(t)=\left(
                   \begin{array}{ccc}
                     \cos\phi _c& -\cos\phi_\tau\sin\phi_c& \sin\phi_\tau\sin\phi_c \\
                     \sin\phi _c& \cos\phi_\tau\cos\phi_c& -\sin\phi_\tau\cos\phi_c \\
                     0 & \sin\phi_\tau & \cos\phi_\tau \\
                   \end{array}
                 \right).
                 \label{evoluB}
\end{equation}
The phase $\phi_\tau(t)$ describes the difference accumulated between those of the two eigenstates
$|\psi_\pm(t)\rangle$ of $I(t)$, i.e.,
$\phi_\tau(t)\equiv\phi_{\tau}^+(t)-\phi_{\tau}^-(t)$ with
$\phi^\pm_\tau(t)=\int_{t_0}^t\langle \psi_\pm(t^\prime)|i\partial_{t^\prime}-H(t^\prime)/\hbar|\psi_\pm(t^\prime)\rangle dt^\prime$.
The complete confinement condition presented in Eq. (\ref{cond}) hence
is equivalent to that all entries of $\mathcal{U}(t)$ should fulfill
$\langle\mathcal{U}_{ij}(t)\rangle_T=0$ as $T\rightarrow \infty$.
The latter can be met in principle if the trigonometric sine and cosine
of $\phi_{c,\tau}(t)$ are periodic and have vanishing integrals
over their corresponding periods.

\begin{figure}
	\includegraphics[width=0.8\columnwidth]{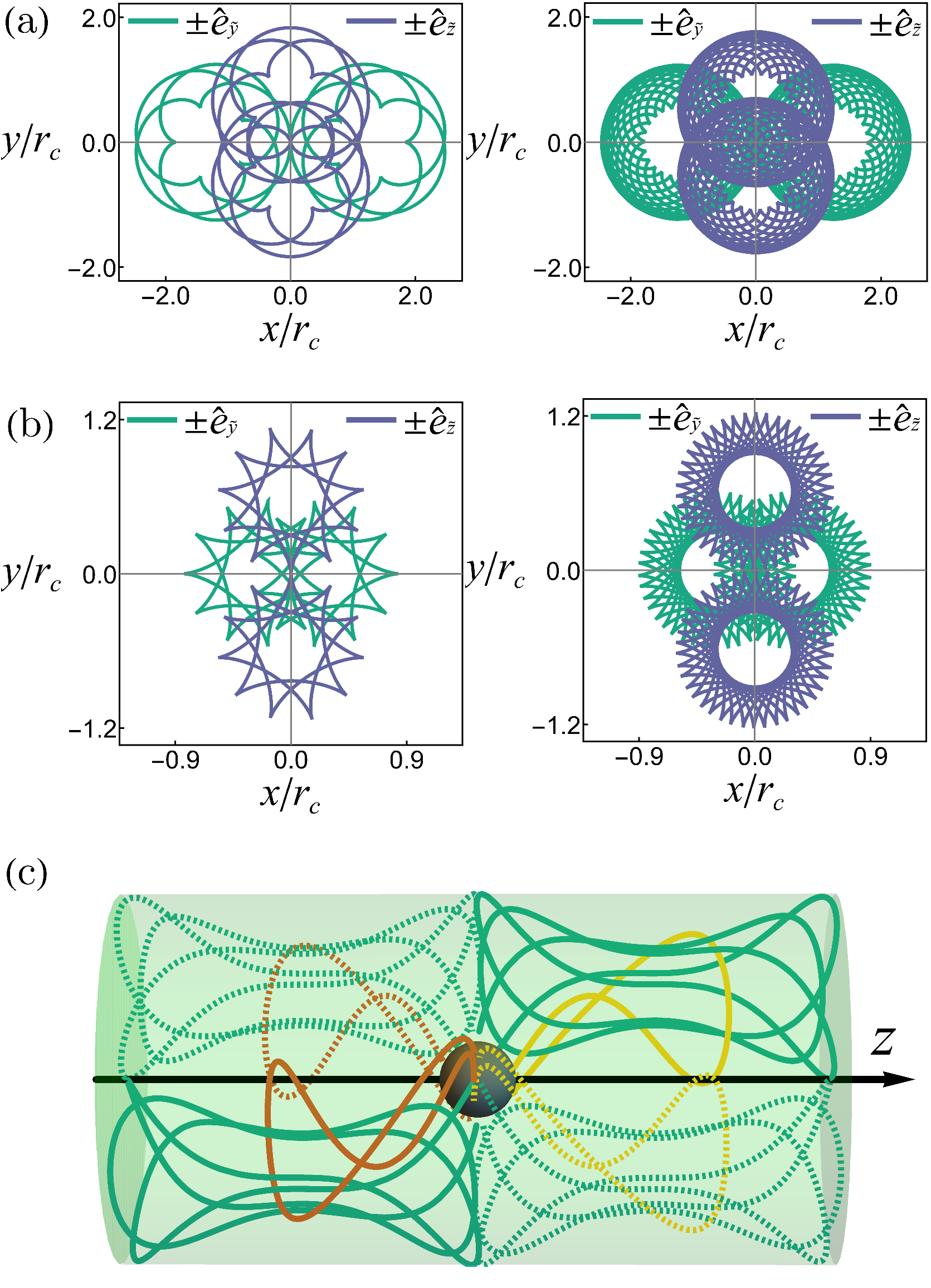}
	\caption{\label{fig1tra} Single-particle motion governed by the pure magnetic force of the RMF
protocol, in which nondimensionalized coordinates, i.e., ratios of $r_i(=x,y,z)$ to the Larmor
radius $r_c\equiv v/\omega_c$, are adopted.
With initial velocities $\vec{v}(0)=v\hat{v}_0$ along $\pm \hat{e}_y$ or $\pm \hat{e}_z$, the
projected trajectories on the $x$-$y$ plane are characterized by similar cycloid curves,
e.g., the green curves ($\hat{v}_0=\pm\hat{e}_y$) are specified by:
$(x-x_1)^2+(y-y_1)^2=\frac {v^2}{(2\omega_+)^2}$, in which
$x_1(t)=\frac {v\cos(\omega_-t)}{2\omega_{-}}+x_0$ and $y_1(t)=\frac {v\sin(\omega_{-}t)}{2\omega_{-}}+y_0$
with
$\omega_\pm\equiv\omega_c\pm\omega_\tau$ and $(x_0,y_0)=(\pm\frac {v\omega_c}{\omega_{+}\omega_{-}},0)$.
(a) Periodic and quasi-periodic epicycloids with $\omega_c/\omega_\tau=2.2$ (left) and
$2.2\pi/3$ (right). (b) Periodic and quasi-periodic hypocycloids
with $\omega_c/\omega_\tau=1/2.2$ (left) and $\pi/6.6$ (right).
(c) Schematic of the trajectory distribution with $\omega_c/\omega_\tau=2.2$.
The solid curves
account for $\hat{v}_0=\pm \hat{e}_y$ (green), $(\hat{e}_x+\hat{e}_y)/\sqrt{2}$ (yellow) and
$-(4\hat{e}_x+3\hat{e}_y)/5$ (orange), respectively, and the axial symmetry of the profile
is indicated by dashed lines.
}
\end{figure}

Let us consider a primitive protocol specified by the rotating-field driven spin model $H(t)$
with $\vec{\Omega}(t)=-\Omega_z\hat{e}_z-\vec{\Omega}_\tau(t)$, where
the transverse component $\vec{\Omega}_\tau(t)=\Omega_\tau[\cos(\omega_ct)\hat{e}_x-\sin(\omega_ct)\hat{e}_y]$
represents a constantly rotating field and the minus is introduced due to the correspondence
$\vec{\Omega}(t)\leftrightarrow -\hbar(q/m)\vec{B}(t)$.
We set $\omega_c\equiv\Omega_z/\hbar$ hence the
dynamical invariant of the system is obtained as \cite{ding2010}
$I(t)=\frac12[\cos(\omega_ct)\sigma_x-\sin(\omega_ct)\sigma_y]$.
Consequently the eigenstates of $I(t)$ reads
$|\psi_\pm(t)\rangle =(e^{-i\omega_ct/2}|0\rangle \pm e^{i\omega_ct/2}|1\rangle)/\sqrt{2}$
and the phases in $\mathcal{U}(t)$ are given by $\phi_c(t)=-\omega_c t$ and $\phi_\tau(t)=\Omega_\tau t/\hbar$.
By applying the product-to-sum formula to the four quadratic entries of $\mathcal{U}(t)$,
it is seen that all $\mathcal{U}_{ij}(t)$ fulfill the condition
$\langle\mathcal{U}_{ij}(t)\rangle_T=0$ in the long-time limit
as long as $\Omega_\tau\neq\Omega_z$.

The confinement property of the corresponding RMF protocol can be clearly
displayed via the orbit motion of the charged classical particle.
The transverse magnetic field of the protocol
$\vec{B}_\tau(t)=B_\tau[\cos(\omega_ct)\hat{e}_x-\sin(\omega_ct)\hat{e}_y]$
rotates (clockwise) at a frequency $\omega_c=qB_z/m$ that is resonant with the ion cyclotron
around the longitudinal $B_z$.
In a reference frame with the velocity rotating
synchronously with $\vec{B}_\tau(t)$, that is,
$\sum_kv_k\hat{e}_k\equiv \sum_kv_k^\prime \hat{e}_k^\prime$ with
$\hat{e}_z^\prime =\hat{e}_z$
and
\begin{equation}
\bigg\{\begin{array}{ll}
  \hat{e}_x^\prime =\cos(\omega_ct)\hat{e}_x-\sin(\omega_ct)\hat{e}_y, \\
  \hat{e}_y^\prime =\sin(\omega_ct)\hat{e}_x+\cos(\omega_ct)\hat{e}_y,
\end{array}
\label{rframe}
\end{equation}
the particle is only affected by the transversal magnetic field $B_\tau$
along the fixed $\hat{e}_x^\prime$. Hence it will either maintain its velocity
parallel to the $\hat{e}_x^\prime$ if initially moving along the $x$ direction
or gyrate at a frequency $\omega_\tau=qB_\tau/m$
around $\hat{e}_x^\prime$ if initially possessing the $v_y$ or $v_z$ component.
That is to say, the Larmor cyclotron ($\omega_c$-cyclotron)
can be regarded as the drifting motion of the ``guiding
center" of the $\omega_\tau$-cyclotron that encircles
around $\vec{B}_\tau(t)$.
In the coordinate space, the yielded longitudinal motion of the trajectory
indicates a standard vibration whereas the transverse one [see Fig. 1(a) and (b)]
gives rise either to a hypocycloid ($\omega_c<\omega_\tau$) or to an
epicycloid ($\omega_c>\omega_\tau$). The general orbit of the particle,
described by the synthesis of the motion patterns through
decomposing $\vec{v}(0)$ into $x$, $y$ and $z$ components, turns out to be
periodic or quasi-periodic in all the three spatial dimensions except for
the particular case of $\omega_c=\omega_\tau$.

The insight gained from the above is that the torus configuration may
not be a necessity when utilizing the TVMF confinement protocol. The particles
emanated stochastically from the centre region of the field distribution will
be blown back by the Lorentz force repetitively [see Fig. 1 (c)],
so the confined system, if ignoring the Coulomb interaction,
would be of solid structure and with a high density in its center.
On the other hand, it is known that the RMF protocol
at the Larmor cyclotron frequency is unable to confine the charged
particle when including
the influence of inductive electric fields \cite{hugrass1983}.
The obstacle, which we will return back to later on, can be attributed to the
resonance accelerating effect of the electric field. The question hence
becomes whether there exists a TVMF protocol that can be robust against
the influence caused by its associated electric field.

Below we present the second TVMF confinement protocol which seems to be
more economic and able to preserve its confinement capability in the presence
of the inductive electric field. The proposal is based on the
(co)tangent-pulse driven quantum model \cite{yang2018} and the driving field herein
takes the form
$\vec{\Omega}(t)=\Omega_x\cot(\omega_ct)\hat{e}_x-\Omega_z\hat{e}_z$ with
$\Omega_z^2=\Omega_x^2+(\hbar\omega_c)^2$.
This model has a dynamical invariant $I(t)=e^{-i\frac{\omega_ct}{2}\hat{e}_{\tilde{z}}\cdot\vec{\sigma}}(\sigma_x/2)e^{i\frac{\omega_ct}{2}\hat{e}_{\tilde{z}}\cdot\vec{\sigma}}$,
in which $\hat{e}_{\tilde{z}}= \cos \varphi \hat{e}_y-\sin\varphi \hat{e}_z$ with
$\varphi=-\arccos(\Omega_x/\Omega_z)$. It evolves along the geodesic loop
normal to the $-\hat{e}_{\tilde{z}}$ direction [see Fig. 2(a)].
In the reference frame described by
$\{\hat{e}_x,\hat{e}_{\tilde{y}}=\sin\varphi \hat{e}_y+\cos\varphi\hat{e}_z,\hat{e}_{\tilde{z}}\}$,
one can obtain that the evolution operator $\mathcal{U}(t)$ takes the same form of
Eq. (\ref{evoluB}) with $\phi_c(t)=-\omega_ct$ and
$\phi_\tau(t)=(\Omega_x/\hbar)\int_{t_0}^t\csc(\omega_ct^\prime)dt^\prime$.

\begin{figure}
	\includegraphics[width=0.8\columnwidth]{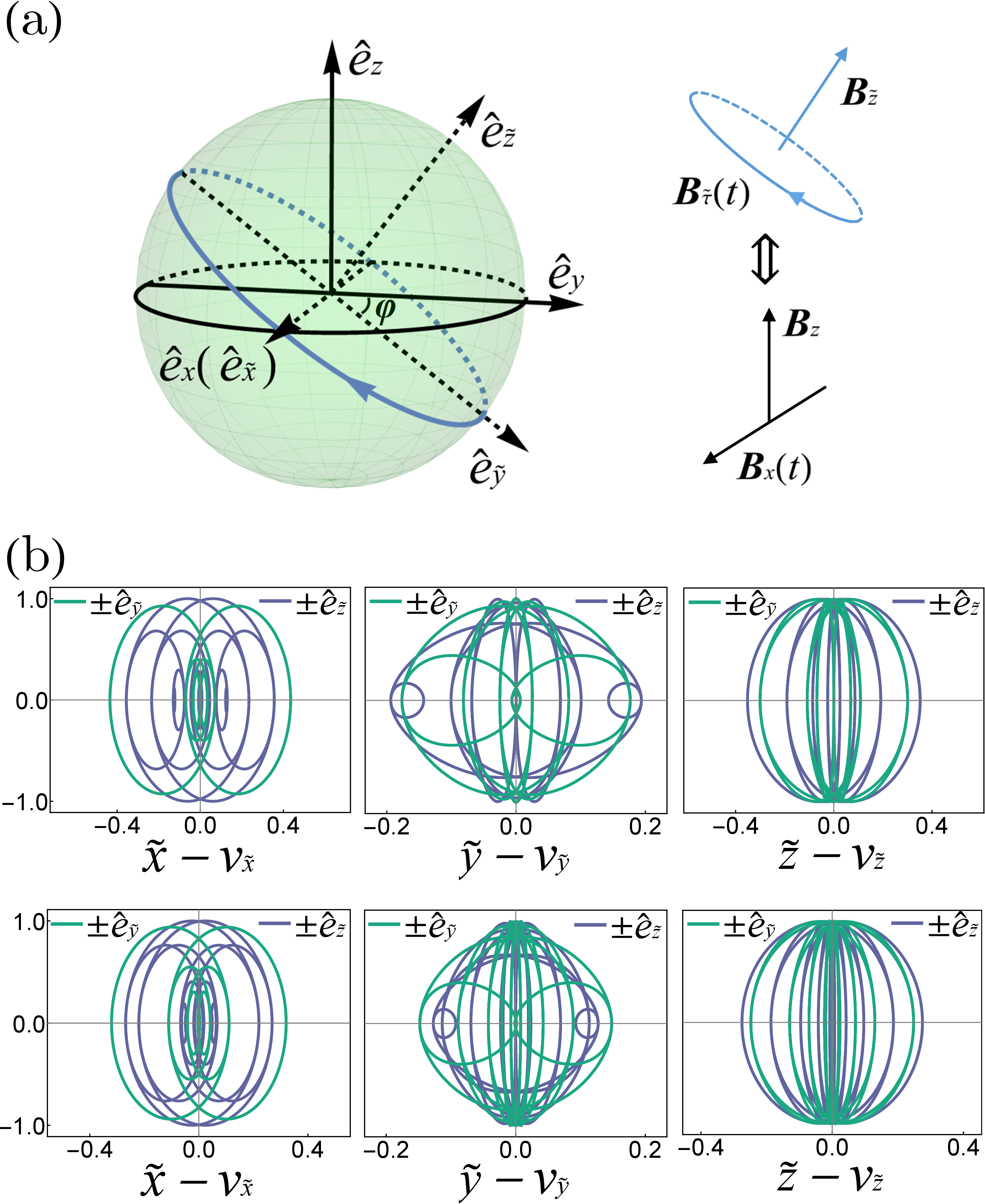}
	\caption{\label{fig2} Schematic diagram of the tangent-pulse protocol
and the bounded orbits governed by the magnetic force. (a) Geodesic-loop evolution of the L-R invariant
$I(t)$ in the Bloch space (left panel) and re-decomposition
of $\vec{B}(t)$ into a ``RMF" form (right panel).
(b) Planar projections of various particle trajectories ($\hat{v}_0=\pm\hat{e}_{\tilde{y}},\pm\hat{e}_{\tilde{z}}$)
in the six-dimensional phase space specified by
$\{\tilde{x},\tilde{y},\tilde{z},v_{\tilde{x}},v_{\tilde{y}},v_{\tilde{z}}\}$,
with field parameters $\{\kappa_1,\delta_1\}$ (upper panel) and
$\{\kappa_2,\delta_2\}$ (lower panel) given in Table I.
Dimensionless $\{\tilde{x}/r_c,\tilde{y}/r_c,\tilde{z}/r_c\}$ ($r_c\equiv v/\omega_c$)
are adopted for the coordinate components
and $\{v_{\tilde{x}}/v,v_{\tilde{y}}/v,v_{\tilde{z}}/v\}$ for the velocity components,
respectively.
}
\end{figure}

This driven quantum model gives rise to a ``linearly polarized" TVMF protocol specified by a constant
$\vec{B}_z=B_z\hat{e}_z$ and a tangent-pulse
transverse field: $\vec{B}_x(t)=B_x\tan(\omega_c t-\frac \pi 2)\hat{e}_x$ with
$\omega_c =(q/m)\sqrt{B^2_z-B^2_x}$. Interestingly, one can re-decompose
$\vec{B}(t)\equiv\vec{B}_x(t)+\vec{B}_z$ into the
coordinate $\{\tilde{x}(=x),\tilde{y},\tilde{z}\}$ as
$\vec{B}(t)=\vec{B}_{\tilde{\tau}}(t)+\vec{B}_{\tilde{z}}$ [see Fig. 2 (b)],
that is, a fixed ``longitudinal" field $B_{\tilde{z}}\equiv(m/q)\omega_c$
along $\hat{e}_{\tilde{z}}$ and a RMF
$\vec{B}_{\tilde{\tau}}(t)=B_{\tilde{\tau}}(t)[\cos(\omega_ct)\hat{e}_x-\sin(\omega_ct)\hat{e}_{\tilde{y}}]$
in the $x$-$\tilde{y}$ plane with $B_{\tilde{\tau}}(t)=-B_x\csc(\omega_ct)$.
In the rotating frame $\{\hat{e}_x^\prime,\hat{e}_{\tilde{y}}^\prime\}$ that
relates $\{\hat{e}_x,\hat{e}_{\tilde{y}}\}$ via Eq. (\ref{rframe}),
the particle is then only affected by the time-varying magnetic field $B_{\tilde{\tau}}(t)$
which is along the fixed $\hat{e}_x^\prime$. That is to say,
the orbit motion is described by a combination of the $\omega_c$-cyclotron
and a frequency-varying gyration governed by the field line $B_{\tilde{\tau}}(t)\hat{e}_x^\prime$.

In the actual process we assume that the periodic transverse pulses
$B_x\cot\phi_c(t)\hat{e}_x$ have symmetric truncations, that is, the angle
``$-\phi_c(t)$" changes from $n\pi+\delta$ to $(n+1)\pi-\delta$ repetitively,
in which $n=0,1,2,\cdots$ and $\delta$ specifies the cutoff of the phase angle.
The accumulated phase $\phi_\tau(t)=(q/m)\int_{t_0}^tB_{\tilde{\tau}}(t^\prime)dt^\prime$
over any two consecutive pulses
will counteract each other due to the antisymmetric
relation: $B_{\tilde{\tau}}(\pi/\omega_c+t)=-B_{\tilde{\tau}}(\pi/\omega_c-t)$.
The periodicity of $\phi_\tau(t)$ and hence that of $\sin\phi_\tau(t)$
and $\cos\phi_\tau(t)$ are naturally guaranteed.
The mean-zero property of $\sin\phi_\tau(t)$ and $\cos\phi_\tau(t)$,
i.e., with vanishing integrals over the pulsing process,
can be achieved through fine-tuning the field parameters
$\kappa\equiv B_x/B_{\tilde{z}}$ and $\delta$. We present two sets of
them in Table I and show the bounded orbits
of the particle in Fig. 2 (b). For the deuteron with a thermal energy of
$20~{\rm KeV}$ (about $2\times 10^8~{\rm Kelvin}$) and the Larmor radius of $20~{\rm cm}$
($\omega_c\approx 5~{\rm MHz}$), the magnetic induction $B_{\tilde{z}}\approx 0.1~{\rm T}$
and the cutoff values of the $\vec{B}_x(t)$ pulse with respect to
$\{\kappa_{1},\delta_{1}\}$ and $\{\kappa_{2},\delta_{2}\}$
are about $4.8~{\rm T}$ and $15.4~{\rm T}$, respectively.

\begin{table}[t]
\caption{Fine-tuned field parameters for the tangent-pulse TVMF protocol.}
\label{twod}
\begin{ruledtabular}
\begin{tabular}{cccc}
$   \{\kappa,~~\delta\}   $    &  $\cot\delta$         &   $B_x/B_z$     & $\phi_\tau(\frac {\pi-\delta}{\omega_c})$  \\ \hline
$	\{\kappa_1,\delta_1\}=\{3.250,0.0678\}	$	 &	$	14.72	$	     &	    $0.9558$    &	$7\pi$	    \\
$	\{\kappa_2,\delta_2\}=\{3.973,0.0258\}	$	 &	$	38.68	$	     &      $0.9698$    &	$11\pi$	 	\\
\end{tabular}
\end{ruledtabular}
\end{table}

The described TVMF confinement protocols till now involve only the magnetic force.
To check if they are valid in the presence of inductive electric fields,
one should resort to the equation of motion
\begin{equation}
\frac {d}{dt}\vec{v}(t)=\frac qm\vec{v}(t)\times \vec{B}(t)+\frac qm\vec{E}(\vec{r},t).
\label{EOMM}
\end{equation}
In general, $\vec{E}(\vec{r},t)$ is determined by the
differential equation
$\nabla\times\vec{E}(\vec{r},t)=-\frac {\partial}{\partial t}\vec{B}(t)$
and the boundary condition. We assume that the associated electric fields
of the RMF and the tangent-pulse magnetic field, denoted respectively
by $\vec{E}_1(\vec{r},t)$ and $\vec{E}_2(\vec{r},t)$, take the form
\begin{equation}
\vec{E}_1(\vec{r},t)=-\omega_cz(t)\vec{B}_\tau(t);
~~\vec{E}_2(\vec{r},t)=\frac {B_x\omega_c\tilde{z}(t)}{\sin^2(\omega_ct)}\hat{e}_{\tilde{y}}.
\label{electr}
\end{equation}
Eq. (\ref{EOMM}) is then resolved numerically with $\vec{r}(t_0)=0$ and
various initial $\vec{v}(t_0)=v_0\hat{v}_0$. The results (see Fig. 3 for two samples)
show that the orbit of the RMF system is unconfined, but the confinement
property of the tangent-pulse protocol is retained, with
the particle shuttling from the centre to a maximum distance
(mostly in the azimuthal direction vertical to the $\tilde{z}$ axis)
$r\approx 30 v_0/\omega_c$.

\begin{figure}
	\includegraphics[width=0.9\columnwidth]{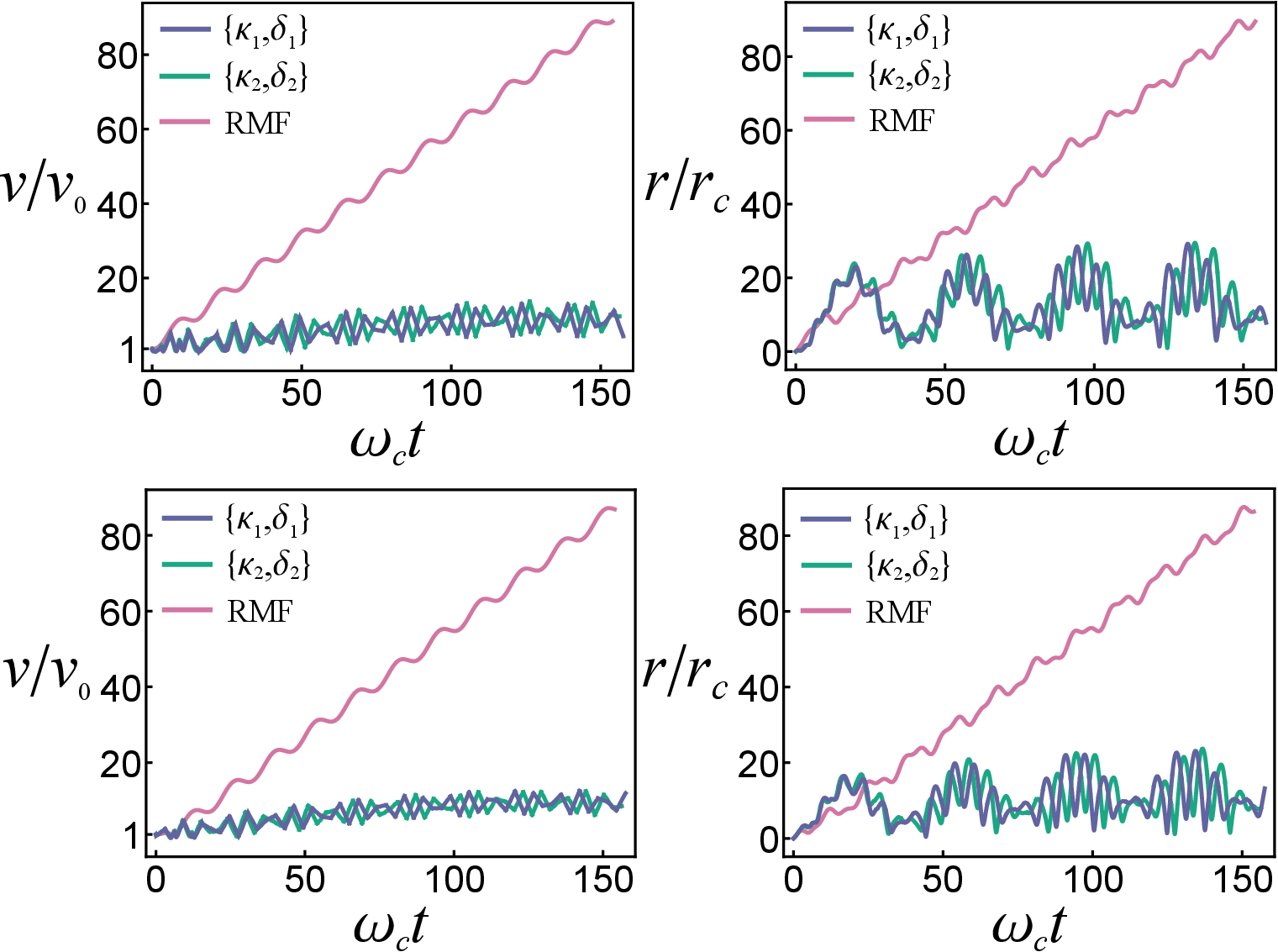}
	\caption{\label{fig1tra} Particle motion within electromagnetic fields
of the RMF protocol (with $\omega_c/\omega_\tau=2.2$) and the tangent-pulse protocol.
The relative magnitudes of the velocity (against the initial speed $v_0$) and
the displacement (against the Larmor radius $r_c=v_0/\omega_c$) have been used.
The initial velocities are along $\hat{v}_0=(1,1,1)/\sqrt{3}$ (upper panel)
and $\hat{v}_0=(1,-1,1)/\sqrt{3}$ (lower panel).
}
\end{figure}

The particle motion within the electromagnetic fields of the RMF protocol is
sensitive to its initial velocity and is unstable
when $\vec{v}(t_0)$ contains the $y$-component. This can be seen since $\vec{E}_1(\vec{r},t)$ rotates
(anti-)synchronously with the RMF in the $x$-$y$ plane,
the produced azimuthal electric force will lead to resonance (de-)acceleration
on the $\omega_c$-cyclotron and the transverse orbit will expand
over time if $\langle z(t)\rangle\neq 0$. The latter comes into existence
as long as $v_y(t_0)\neq0$ [the longitudinal motion is not affected
by $\vec{E}_1(\vec{r},t)$, c.f. Fig. 1(c)].
By contrast, the electric field $\vec{E}_2(\vec{r},t)$ excited in the
tangent-pulse protocol is linearly polarized so the resonance acceleration
can be avoided. The combination of the chirped magnetic pulses and electric
pulses offers a stable confinement regime which is insensitive to the initial
state of motion of the particle. The latter property has been confirmed
for randomly sampled orientations of $\vec{v}(t_0)$ by numerically resolving
the orbits over durations up to several thousands gyroperiods.

The confined single-particle motion under the TVMF protocol suggests a
possible confinement scenario for fusion plasmas which may offer
advantages such as a high reaction rate and increased tolerance to plasma instability
owing to its centralized configuration.
Note that the charge-mass ratio of the fuel triton is
two-thirds of that of the deuteron. To effectively confine both of these ion species,
a ``magnetic bottle" structure with layered magnetic induction intensity
[e.g., $B_{x,z}^{(t)}=1.5B_{x,z}$ at two side layers for confining
the triton]
should be exploited. Since the Larmor frequency of electrons is far from
synchronous with the field frequency $\omega_c$, they will be poorly confined in the
longitudinal direction until the electric potential becomes strong enough
to provide the confinement. The built-in ambipolar field and the resulting Langmuir
oscillations \cite{langmuir1,langmuir2} in these systems may require more attention compared to
that given to toroidal plasmas.

To summarize, we have demonstrated the potential of achieving a non-torus confinement
configuration for charged particles using the TVMF protocol. The discovery, along with the strategy
we have employed, may open up an alternative way to design the magnetic confinement scenario
for fusion plasmas. Besides, further investigation on the property of the confinement regime,
including the integrability of the nonautonomous dynamical system with respect to
periodically pulsing electromagnetic fields, should be a subject of future study.

\acknowledgments {This work was supported by the NSFC, China, under Grant No. 12147207.}


\end{document}